\documentclass[structabstract]{aa}  

\usepackage{graphicx,epsfig,fancyhdr,rotating,amsmath,natbib,amssymb}

\usepackage{txfonts,graphicx,fixltx2e}

\begin{document}

\title{Observations of dissipation of slow magneto-acoustic waves in a polar coronal hole}

\author{G.~R.~Gupta
}

\institute{Inter-University Centre for Astronomy and Astrophysics (IUCAA), Post Bag-4, Ganeshkhind, Pune 411007, India. \\
	  \email{girjesh@iucaa.ernet.in}
}

\date{Received 5 December 2013 / Accepted 2 July 2014}

\abstract
{}
{We focus on a polar coronal hole region to find any evidence of dissipation of propagating slow magneto-acoustic waves.}
{We obtained time-distance and frequency-distance maps along the plume structure in a polar coronal hole. We also obtained
Fourier power maps of the polar coronal hole in different frequency ranges in 171~\AA\ and 193~\AA\ passbands. We performed
intensity distribution statistics in time domain at several locations in the polar coronal hole.}
{We find the presence of propagating slow magneto-acoustic waves having temperature dependent propagation speeds. 
The wavelet analysis and Fourier power maps of the polar coronal hole show that low-frequency waves are travelling 
longer distances (longer detection length) as compared to high-frequency waves. 
We found two distinct dissipation length scales of wave amplitude decay at two different height ranges (between 0--10 Mm and
 10--70 Mm) along the observed plume structure. The dissipation lengths obtained at higher height range show some frequency dependence. 
Individual Fourier power spectrum at several locations show a power-law distribution with frequency whereas
 probability density function (PDF) of intensity fluctuations in time show nearly Gaussian distributions.}
{Propagating slow magneto-acoustic waves are getting heavily damped (small dissipation lengths) within the first 10~Mm distance.
Beyond that waves are getting damped slowly with height. Frequency dependent dissipation lengths of wave propagation at
higher heights may indicate the possibility of wave dissipation due to thermal conduction, however, the contribution from other dissipative
parameters cannot be ruled out.
Power-law distributed power spectra were also found at lower heights in the solar corona, which may
 provide viable information on the generation of longer period waves in the solar atmosphere.}


\titlerunning{Observations of dissipation of slow magneto-acoustic waves}
\authorrunning{G.~R.~Gupta}

\maketitle

\section{Introduction}

Propagating intensity disturbances in polar coronal holes were detected with the 
UltraViolet Coronagraph Spectrometer (UVCS) on-board the Solar and Heliospheric Observatory (SOHO)
 \citep{1997ApJ...491L.111O} and
 Extreme ultraviolet Imaging Telescope (EIT/SOHO) \citep{1998ApJ...501L.217D} with periods between 10~min to 20~min. 
 These disturbances were 
interpreted in terms of propagating slow magneto-acoustic waves \citep{1999ApJ...514..441O}. 
\citet{2006A&A...452.1059O,2007A&A...463..713O} found evidence of propagating slow magneto-acoustic waves using
 spectroscopic data from  the Coronal Diagnostic Spectrometer (CDS/SOHO), 
 and \citet{2009A&A...493..251G} found similar results from the 
 Solar Ultraviolet Measurement of Emitted Radiation  (SUMER/SOHO) data using
 statistical techniques. Recent spectroscopic observations with the 
EUV imaging spectrometer (EIS/\textit{Hinode})  \citep{2009A&A...499L..29B,2010ApJ...718...11G},
and imaging observation with the Atmospheric Imaging Assembly (AIA) on-board the Solar Dynamics Observatory (SDO)
 \citep{2011A&A...528L...4K}  also detected propagating waves in the polar coronal holes. 
\citet{2011SSRv..158..267B} provides an overview of observational evidences of propagating
magnetohydrodynamics (MHD) waves
 in the coronal hole regions. Similar propagating slow magneto-acoustic waves were also observed in
coronal loops from imaging data obtained from EIT/SOHO \citep{1999SoPh..186..207B} and
Transition Region And Coronal Explorer (TRACE) \citep{2000A&A...355L..23D}
with periods around 5~min. Recent spectroscopic observations with EIS/\textit{Hinode}
\citep{2009A&A...503L..25W} and imaging observations with AIA/SDO \citep{2012SoPh..279..427K,2012A&A...543A...9Y}
 also show similar propagating waves in the coronal loops. \citet{2009ApJ...697.1674M} obtained stereoscopic
 observations from the Extreme UltraViolet Imager (EUVI) on-board  the Solar TErrestrial RElations Observatory (STEREO)
 and presented a three-dimensional propagation of slow magneto-acoustic waves within
 active region coronal loops. At present, excitation mechanisms for longer period waves in the solar corona are
 still unclear. \citet{2009SSRv..149...65D} and \citet{2012RSPTA.370.3193D} provide comprehensive reviews
 on propagating intensity disturbances along the coronal loops. 

Slow magneto-acoustic waves will be subjected to damping while propagating in the solar atmosphere. 
\citet{2000ApJ...533.1071O} found that slow waves non-linearly steepen while propagating into corona, which leads
  to the enhanced dissipation of the waves. Propagation and damping of slow magneto-acoustic waves in the solar atmosphere
 are very well studied theoretically. Major components considered for damping were compressive viscosity 
\citep{2000ApJ...533.1071O}, thermal conduction \citep{2003A&A...408..755D}, gravitational stratification and field
 line divergence \citep{2004A&A...415..705D}, mode coupling \citep{2004A&A...425..741D}, and shocks 
\citep{2008ApJ...685.1286V}. Thermal conduction and field line divergence appear to be the dominant
 damping parameters for typical coronal conditions \citep{2004A&A...415..705D}, however, as already known,
field line divergence is just  geometrical effect that does not involve any wave dissipation mechanisms. 

\begin{figure*}[htba]
\centering
\includegraphics[width=18cm]{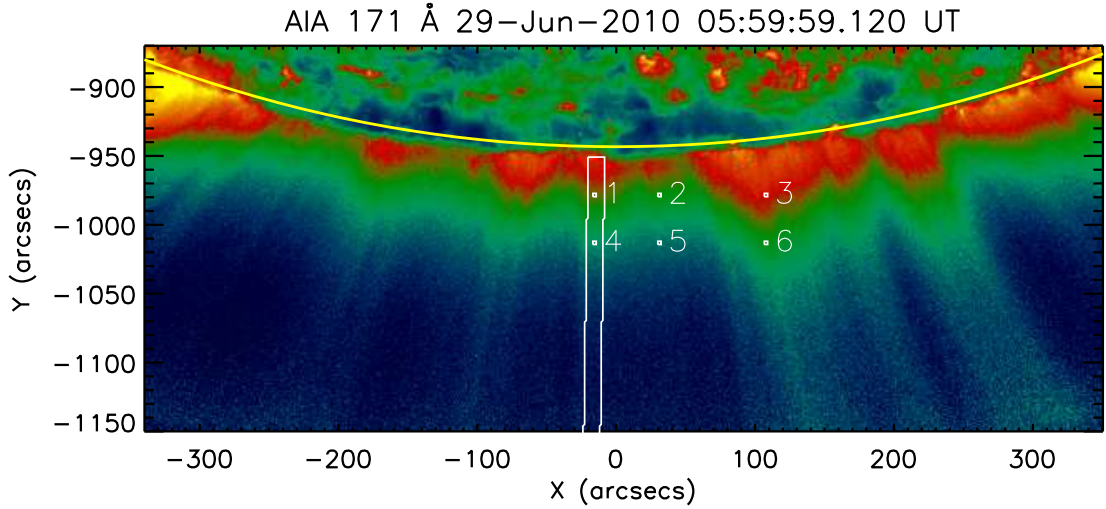}
\caption{Context map of a south polar coronal hole from AIA/SDO 171~\AA\ passband. The continuous white line box in
 off-limb region indicates the selected plume region for which a time--distance intensity map (see Fig.~\ref{fig:int_xt})
 and a frequency--distance Fourier power map (see Fig.~\ref{fig:pwr_xt}) are obtained. Different small
boxes indicate the individual spatial locations chosen for detailed analysis. To make the far off-limb structures
 visible, a radial enhancement filter is applied on the image.}
 \label{fig:context}
\end{figure*}

In astronomy, a collection of unresolved motions in space or time, or dynamical oscillations having a broad-band
 spectrum of frequencies without any dominant frequency, are termed as turbulence \citep{2007AIPC..932..327C}.
In turbulence, the spectrum of variable fields are known to obey a power-law distribution over a range of scales
\citep{1941DoSSR..30..301K,1964SvA.....7..566I,1965PhFl....8.1385K}. Turbulence is ubiquitous in the Sun and
 heliosphere and power-law distributions are observed in the solar atmosphere, solar wind, and interplanetary medium
 \citep[detailed in ][]{2005LRSP....2....4B,2010SSRv..156..135P}. Turbulence can play an important role in
wave excitation. \citet{2001A&A...370..136S} developed theoretical formalism of excitation for stellar p-modes
 by turbulent convection and applied it to the Sun \citep{2001A&A...370..147S}.
Moreover, \citet{2004ApJ...604..671F} indicated the importance of background turbulence in the damping
of MHD waves. Thus, turbulence can be considered as an important process  to study excitation and damping
mechanisms of MHD waves in the upper solar atmosphere.    

In this paper, we focus on polar coronal holes, which are considered as a source regions of high-speed
 solar wind streams \citep{1973SoPh...29..505K}. In this region, field lines are open and almost radial, 
which allows us to trace them easily for detailed study. For more details about the region, see the review by
\citet{2009LRSP....6....3C}. We demonstrate the presence of propagating waves along the coronal structures in
the off-limb region (sec.~\ref{sec:xt}). We create Fourier power maps in different frequency ranges of the 
coronal hole (Sec.~\ref{sec:pwr}) and choose one plume structure for detailed study (Sec.~\ref{sec:pwr_xt}).
We investigate the dissipation of propagating waves after taking the effects of gravitational stratification
 and field line divergence into account (Sec.~\ref{sec:damping}). Further, we discuss the origin and source of various results
presented in the previous sections (Sec.~\ref{sec:turbulence}), and, finally, a summary of our results are presented
 (Sec.~\ref{sec:conclusion}). 
       
\section{Observations}

AIA/SDO provides full-disk solar images in 
seven EUV and three UV-visible channels covering a temperature range from $6\times10^4$~K to $2\times10^7$~K. 
The instrument records continuous images of the Sun with spatial resolution of $1.5\arcsec$\ and temporal resolution of 12~s
\citep{2012SoPh..275...17L}. We chose a three hrs time sequence in 171~\AA\ and 191~\AA\ passbands on 2010 June 29 between
 06:00 UT to 09:00 UT from AIA/SDO in the south polar coronal hole. Emission in these two passbands mainly comes from 
Fe~{\sc ix/x} and Fe~{\sc xii} ions. All the images were calibrated, co-aligned, and rescaled  to a
 common $0.6\arcsec$\ plate scale using  $aia\_prep.pro$  routine available in the SolarSoft (SSW).
All the images were rebinned over $2\times2$ spatial pixels to improve the signal strength in the off-limb region.
 Fig.~\ref{fig:context} shows the region of the south polar coronal hole analysed in this study. A radial filter was applied
 to the image to enhance the visibility of coronal structures in the far off-limb region.

\section{Results and discussion}
\label{sec:result}
\subsection{Propagating disturbances in the polar coronal hole}
\label{sec:xt}
\begin{figure}[htbp]
\centering
\includegraphics[width=7.5cm]{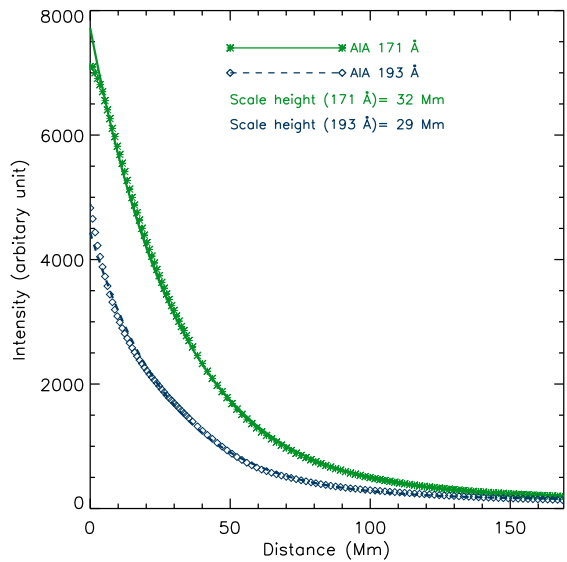}
\caption{Average intensity decrease with height along the plume structure marked in Fig.~\ref{fig:context}. Intensity decrease
are fitted with exponential function to obtain the intensity decay scale heights in AIA 171~\AA\ and 193~\AA\ passbands
 and are labelled in the figure.}
 \label{fig:scale_ht}
\end{figure}

In the south polar coronal hole, several coronal structures such as plumes and inter-plumes can be identified
 from the intensity contrast (see Fig.~\ref{fig:context}). 
At these locations, artificial slits were traced from the near limb region to close to the 
end of the AIA/SDO field of view (FOV). The width of the slits were kept about $12\arcsec$ in the solar-$X$ direction. 
Time--distance maps were created using the intensity along these slits. 
One such plume structure marked in Fig.~\ref{fig:context} was chosen for further detailed study. 
Average intensity decrease along the plume structure were obtained in AIA 171~\AA\ and 193~\AA\ passbands and fitted with
exponential function $I=I_0 exp(-h/H_I)+c$ using the MPFIT routines \citep{2009ASPC..411..251M}
 to obtain the intensity decay scale heights ($H_I$) in both the passbands (see Fig.~\ref{fig:scale_ht}).
 The obtained intensity scale heights were 32~Mm and 29~Mm for AIA 171~\AA\ and
 193~\AA\ passbands, which leads to the density scale height to be 64~Mm and 58~Mm for both the
passbands as $I\propto \rho ^2$.

\begin{figure}[htbp]
\centering
\includegraphics[width=9cm]{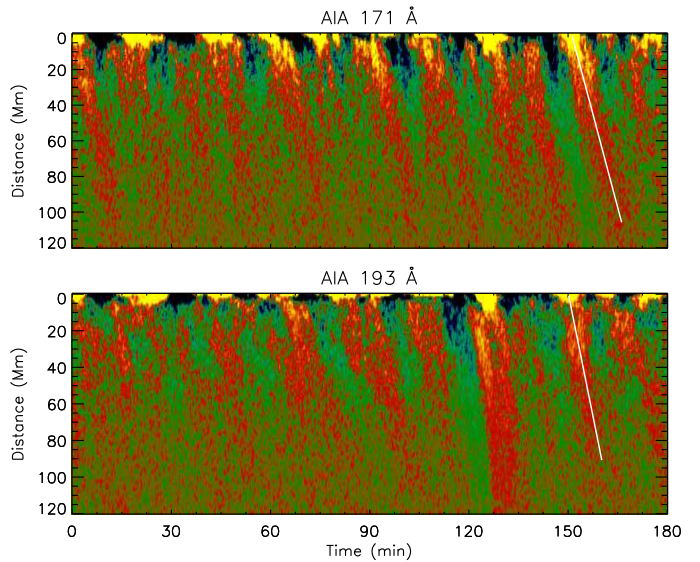}
\caption{Processed time--distance intensity maps of plume region marked in Fig.~\ref{fig:context} summed over
$12\arcsec$ width in the $X$-direction. The quasi-periodic (in the range 10--20~min)
propagating disturbances are observed with speed $\approx 116.5\pm 4.6$~km~s$^{-1}$ in 171~\AA\ (\textit{top panel})
and  $\approx 149.7\pm 6.4$~km~s$^{-1}$ in 193~\AA\ (\textit{bottom panel}) AIA/SDO channels. 
Disturbances are reaching very far in the off-limb corona.}
 \label{fig:int_xt}
\end{figure}

In Fig.~\ref{fig:int_xt}, we show the presence of propagating disturbances from the time--distance map along the plume
 structure obtained in AIA 171~\AA\ and 193~\AA\ passbands. Time--distance maps are created after subtracting
  100-point running average in time ($\approx 20$~min) from each spatial pixels along the artificial slit. 
The slope of propagating ridges provide a speed of propagation that were measured about $116.5\pm 4.6$~km~s$^{-1}$
and $149.7\pm 6.4$~km~s$^{-1}$ in 171~\AA\ and 193~\AA\ passbands.
 Almost all the structures show the presence of propagating disturbances reaching very far out in the corona with the
 dominant period range between 10 min to 25 min, as also reported by \citet{2011A&A...528L...4K} in different AIA channels.
Measured propagation speeds are temperature dependent. The ratio of propagation speed between 193~\AA\ and 171~\AA\ channels
is about 1.29 whereas the ratio of the peak temperature of dominant contributing ions in the two filters can be anywhere
 in the range 1--1.58--1.99--2.24 in a coronal hole \citep[dominant ion contributing in 171~\AA\ channel is Fe~{\sc ix} whereas
that in 193~\AA\ channel are Fe~{\sc ix}, Fe~{\sc x}, Fe~{\sc xi}, and Fe~{\sc xii},][]{2010A&A...521A..21O}.
These two ratios are close enough to follow the relation of acoustic waves $C_s\propto T^{1/2}$ \citep{1984smh..book.....P}.    
 \citet{2009A&A...499L..29B}  interpreted propagating disturbances in the coronal hole in terms of slow
magneto-acoustic  waves based on their temperature dependent propagation speeds from the spectroscopic data. 
Recently, \citet{2011A&A...528L...4K} and \citet{2012SoPh..279..427K} also found similar temperature dependent
 propagation speeds of intensity disturbances from AIA/SDO imaging observations and again interpreted them in terms of
slow magneto-acoustic waves. \citet{2013ApJ...778...26U} developed a new data analysis methodology 
to measure the speed of propagating disturbances. They found that the speed of these disturbances increases with
 the temperature and follows the square-root dependence as predicted for propagating slow magneto-acoustic waves.
\citet{2012A&A...546A..93G} performed a detail line profile study of propagating
 disturbances in a polar coronal hole and found them to be consistent with propagating slow magneto-acoustic waves.
 Henceforth, propagating disturbances observed in the coronal hole with similar properties
 in this study can also be attributed to the presence of propagating slow magneto-acoustic waves.
 
\begin{figure*}[htbp]
\centering
\includegraphics[width=7cm, angle=90]{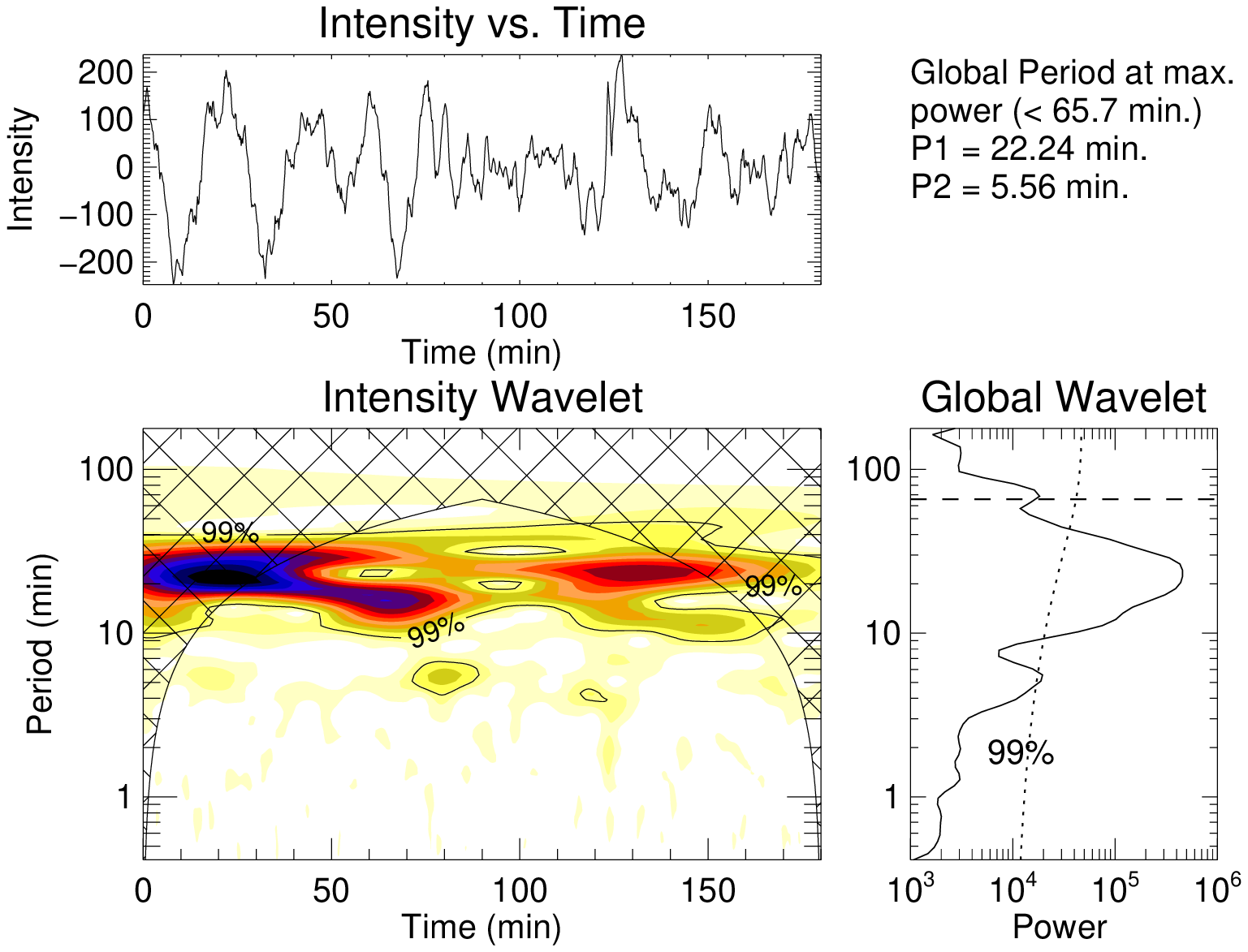}\includegraphics[width=7cm, angle=90]{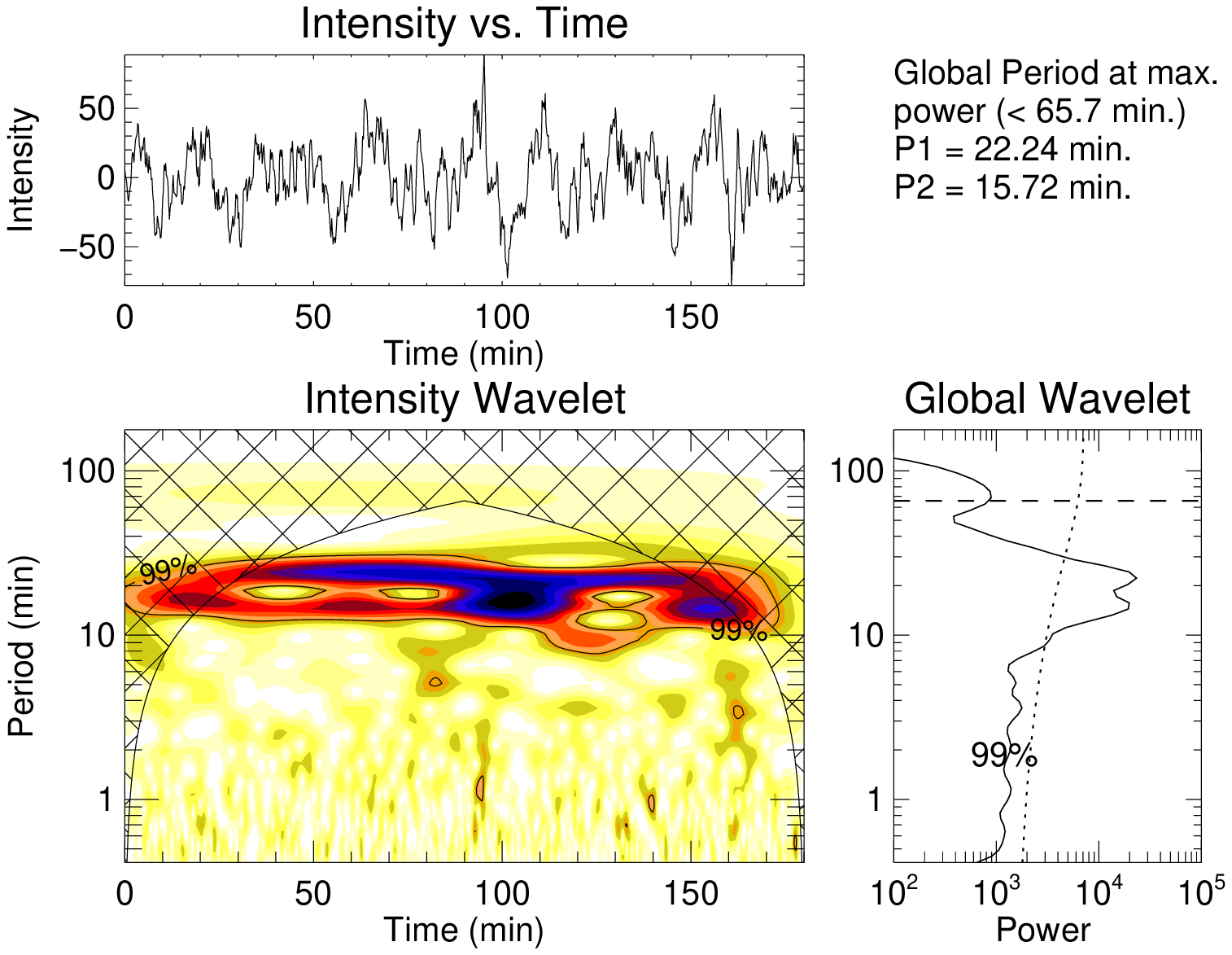}
\caption{The wavelet results for the locations at a distance of 2.6 Mm (left panels) and 20.2 Mm (right panels) 
 from the limb brightening height for a plume region marked in Fig.~\ref{fig:context}  in AIA 171~\AA\ channels.   
 In each set, the top panels show the trend subtracted (100-point running average) intensity variation with time.
 The bottom-left panels show the colour-inverted wavelet power spectrum with 99 \% confidence-level contours,
 while the bottom-right panels show the global wavelet power spectrum (wavelet power spectrum averaged over time)
 with 99 \% global confidence level drawn. The periods P1 and P2 at the locations of the first two maxima
 in the global wavelet spectrum are shown above the global wavelet spectrum.}
 \label{fig:wvlt}
\end{figure*}

\subsection{Fourier power maps of the polar coronal hole}
\label{sec:pwr}

The result presented in Sec.~\ref{sec:xt} provides evidence of  propagating slow magneto-acoustic waves, 
which reach very far out in the off-limb corona. Upon performing the wavelet analysis 
\citep{1998BAMS...79...61T} at two different heights (2.6 Mm and 20.2 Mm from the limb brightening region,
 Fig.~\ref{fig:wvlt}), we found clear presence of both small ($\approx5$~min) and long period ($\approx20$~min)
 oscillations at lower heights whereas at larger heights we found only long period ($\approx20$~min) oscillations.
 This may indicate some period or frequency dependent nature of detection length of wave propagation
 in the solar atmosphere, where detection length is defined as length over which oscillations are visible above
certain noise level \citep{2002SoPh..209...61D}.
Thus, in this section, we look for distance travelled (detection length) by observed propagating waves in 
the polar coronal hole in different period/frequency ranges. To achieve this in detail and for a wide
 range of frequencies, we obtained Fourier power for all the spatial pixels in the polar coronal hole. 
 While applying the Fourier transform on each spatial pixel,
 we subtracted the mean value from the original time series curves without any further trend subtraction or filtration
so as to avoid any kind of bias effect. Thus, a 3-D datacube of Fourier powers were obtained in the polar coronal hole.
While analysing the Fourier powers in off-limb corona, we need to consider the effect of fall off in brightness
 with height. Brightness depends upon density, which decreases with height due to gravitational stratification 
 in the off-limb corona. Wave propagation in the off-limb corona will perturb the density residing on the top of 
density fall. The Fourier power of an oscillation at a particular frequency completely depends upon
 its oscillation amplitude irrespective of the mean value on which it oscillates. Thus, the Fourier power of an
 oscillation with absolute amplitude between any $\pm n$ will always be the same irrespective of the oscillation sitting
 on the top of any mean intensity. Therefore, beacause of fall off in brightness with height, Fourier powers of oscillations
 will remain constant if absolute amplitude of oscillations are constant with height.

\begin{figure*}[htbp]
\centering
\includegraphics[width=17.5cm]{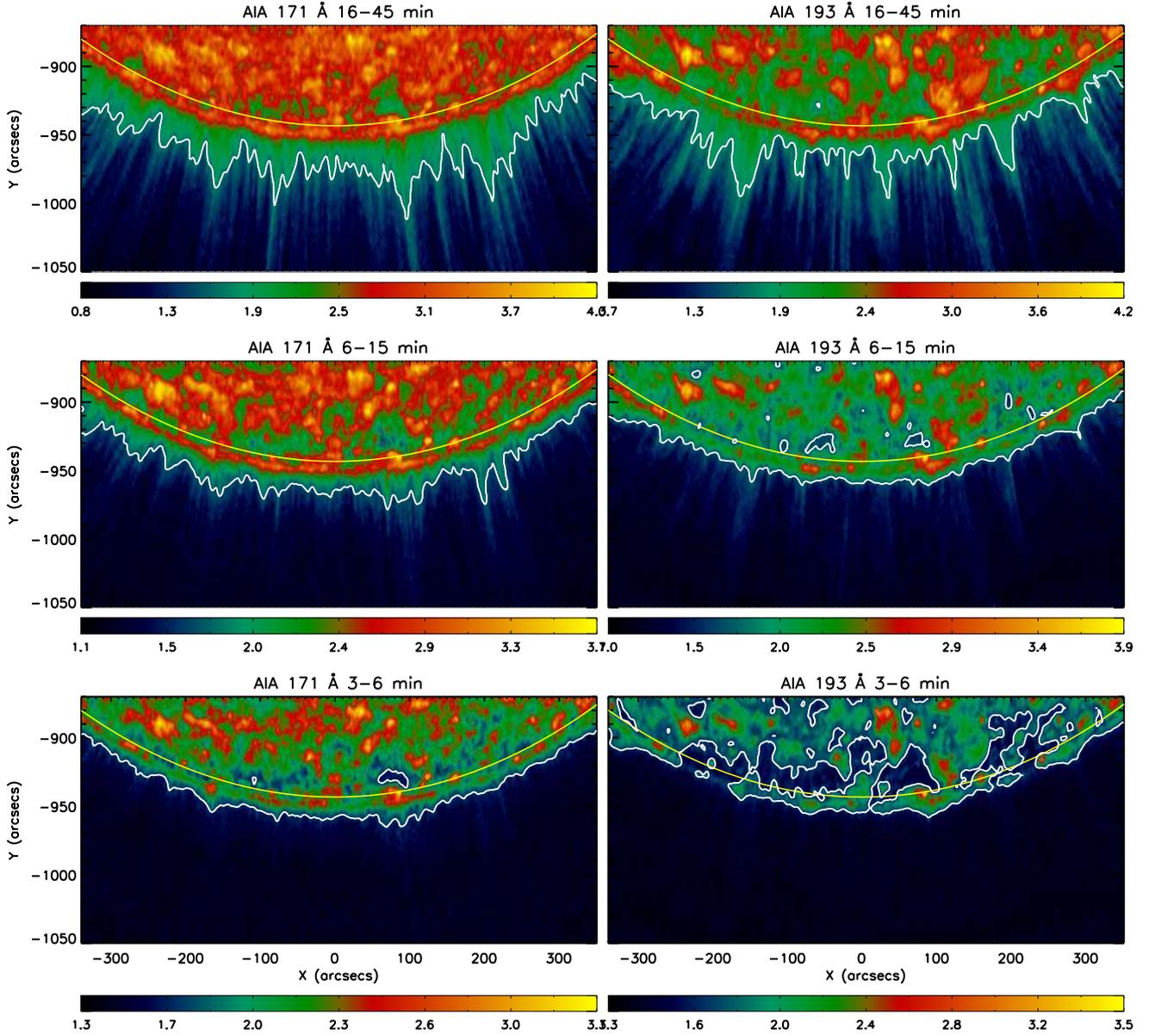}
\caption{The Fourier power map (power in logarithmic scale) of south polar coronal hole obtained from 171~\AA\ (left panels) and 193~\AA\ 
passbands (right panels) of AIA/SDO in different period ranges as labelled. White line contours are plotted at the same
Fourier power level ($\approx$1.7 in log scale) in all the panels of 171~\AA\ and 193~\AA\ passbands. Location
of solar limb is also over-plotted with continuous yellow line in all the panels.}
\label{fig:pwr_map}
\end{figure*}

 We created several frequency windows (ranges) to analyse the 2-D Fourier power map of the polar coronal hole in several 
frequency bins. We divided the frequency scale in three parts to study (i) long periods (periods between 16 min to 45 min),
 (ii) intermediate periods (periods between 6 min to 15 min), and (iii) short periods (periods between 3 min to 6 min).
 We summed the Fourier powers in the respective frequency bins and obtained the Fourier power maps
 of the polar coronal hole in those frequency/period ranges and plotted them in  Fig.~\ref{fig:pwr_map}.

Fig.~\ref{fig:pwr_map} clearly shows that long-period waves reach very large distances in the corona whereas
 short-period waves reach very low distances in the corona for the same Fourier power level.  At first glance, this can be interpreted as
a clear indication of frequency dependent detection length of these waves. Longer period waves have longer
detection length as compared to short period waves, however, one needs to examine
the nature of damping length \citep[distance over which the wave amplitude has an e-folding
 decay, ][]{2002SoPh..209...61D} with frequency (period) before arriving at any conclusion.
 More analysis on damping length of these waves are presented in the next sections. Moreover,
Fourier power spectra at individual spatial locations also show the presence of power peaks at different 
frequencies and confirm the presence of oscillatory signatures as seen in Figs.~\ref{fig:int_xt} and \ref{fig:wvlt}.
 Details about the individual Fourier power spectra are discussed later in Sec.~\ref{sec:turbulence}.   
 From these power maps, it is also noted that 171~\AA\ power contours  provide  finer details of
the coronal structures whereas 193~\AA\ power contours are relatively wider in width. This could be a result of the
 wider temperature response of 193~\AA\ filter in the coronal hole region, as reported by \citet{2010A&A...521A..21O}.   

\subsection{Frequency--distance Fourier power map of propagating disturbances}
\label{sec:pwr_xt}

\begin{figure}[htbp]
\centering
\includegraphics[width=8.5cm]{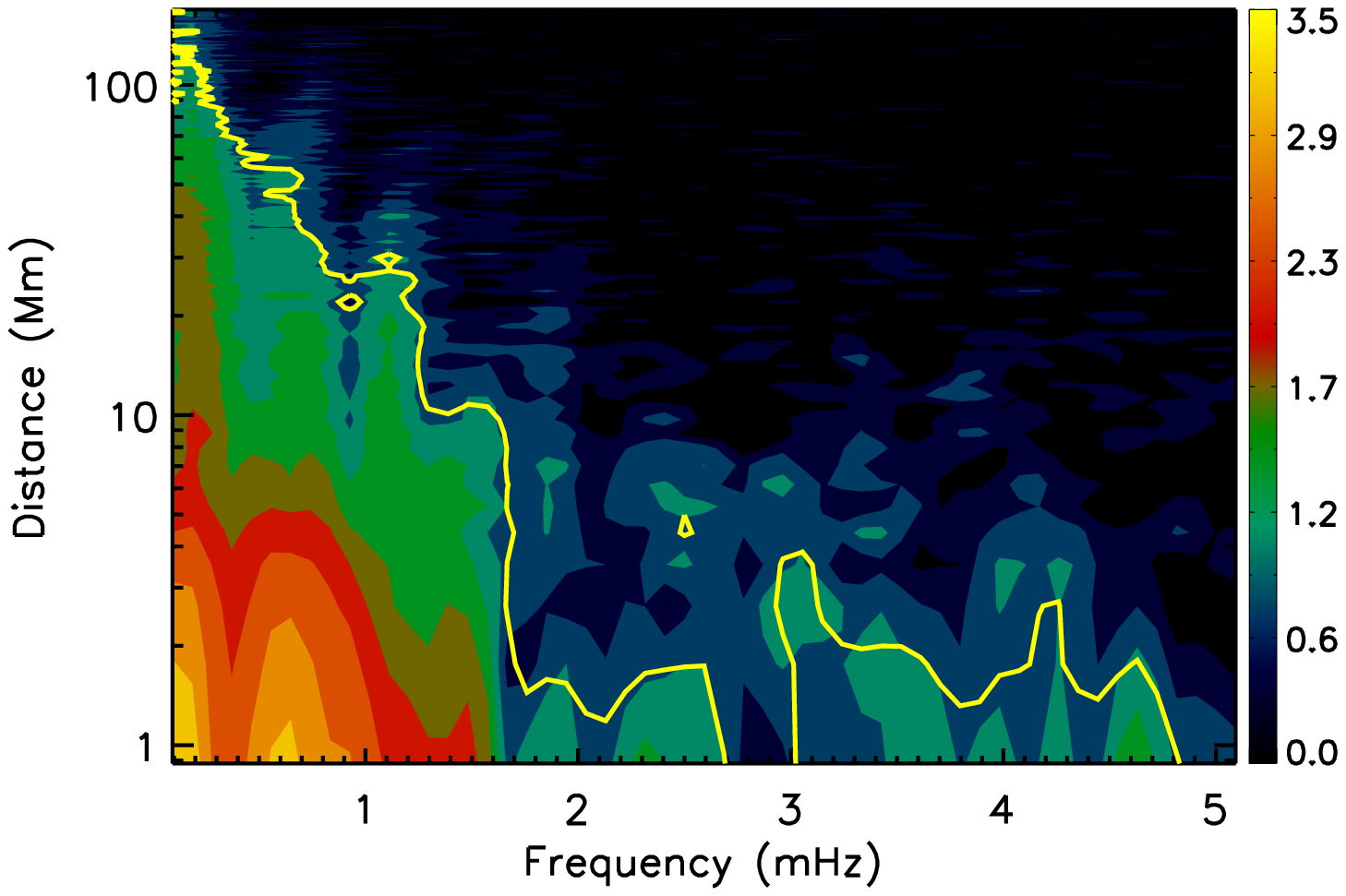}
\caption{Frequency--Distance Fourier power map (power in logarithmic scale) for plume region marked in Fig.~\ref{fig:context} summed over 
$12\arcsec$ width in  $X$-direction. The over-plotted continuous yellow line contour indicates the same Fourier 
power level ($\approx$0.99 in log scale) with frequency.  $Y$-axis (distance axis) is in log scale.}
\label{fig:pwr_xt}
\end{figure}

In order to study the distance travelled (detection length) by waves with different 
frequencies (periods), we analysed the same plume structure selected for the  time--distance map 
in Fig.~\ref{fig:int_xt} and marked with a pair of continuous straight
 lines in the off-limb region of Fig.~\ref{fig:context}.     
In Fig.~\ref{fig:pwr_xt}, we show the Fourier power distribution with frequency and distance above the limb of the
analysed plume region. An isocontour (continuous yellow line) of Fourier power is over-plotted to show the distance
 travelled by waves at different frequencies.
The nature of contour indicates that the decrease of detection length is very rapid with respect to frequency. 
Power peaks observed near 1~mHz ($\approx$17~min) and between 3--5~mHz ($\approx$3--6~min) frequency
 ranges show different length scales of distance travelled. This rapid decrease may explain the detection
 of only longer period waves in the far off-limb corona with EIT/SOHO  \citep[e.g.][]{1998ApJ...501L.217D} 
and UVCS/SOHO \citep{1997ApJ...491L.111O} instruments, and studies further far in the heliosphere
 \citep[see review by ][]{2011SSRv..158..267B}.  The observed detection length of oscillations 
with respect to frequency can be visualised from the contour plot in Fig.~\ref{fig:pwr_xt}.

Observed periods found here ranges from a few minutes to several minutes. 
The acoustic cut-off frequency for the corona with a temperature of 1~MK  is about $70$~min \citep{1999ApJ...514..441O}. 
As periods of oscillations analysed in this study are clearly below 70 min,  if these waves were generated in the low corona,
 they could easily propagate along the field lines and detected up to far distances as found in this study. 
 There are recent reports, however,  providing evidence of longer period 
waves also present at lower solar atmospheric heights of the quiet-Sun network regions \citep{2013SoPh..282...67G}.
Thus, the possibility of the origin of longer period waves observed in the corona to be at lower solar atmospheric heights cannot be ruled out.
In those case, however, the longer period waves above the acoustic cut-off of the respective layer of solar atmosphere will be able to propagate further
only along the inclined field lines. These longer period waves  could be the result of some longer time scale photospheric motions, which
may be verified by observing the spectrum of photospheric motions below the plume structures as suggested by \citet{1999ApJ...514..441O}.

\subsection{Effect of area divergence, gravitational stratification and other parameters}
\label{sec:damping}
In the Sec.~\ref{sec:pwr} and \ref{sec:pwr_xt}, we showed the possible frequency dependent nature of
detection lengths of propagating slow magneto-acoustic waves in the solar corona. However, 
effect of various coronal parameters should be taken into
 account to conclude on any frequency dependent nature of damping lengths. 

As mentioned earlier, Fourier power at a frequency depends upon the absolute amplitude of oscillation
rather than on the relative amplitude. From the Figs.~\ref{fig:pwr_map} and \ref{fig:pwr_xt}, we can see that 
the Fourier powers in different frequency ranges decrease with height, and thus, indicate that the amplitude of waves
 are decreasing with height. This decrease in amplitudes are the result of damping of waves in the solar corona
 due to various mechanisms.
 
In the solar atmosphere, because of gravitational stratification and magnetic field divergence, absolute wave amplitude of density perturbation
 decreases while the relative wave amplitude increases with height in the off-limb corona because of decrease in the background density
  \citep{1999ApJ...514..441O}. This implies that Fourier transform (FT) value at a frequency due to wave propagation, in 
principle, will decrease with height. 
However, one should be careful that decreases in the FT values with height due to stratification and area divergence  
are not actual dissipation of waves as there is no dissipative mechanisms involved. These are purely a 
geometrical effect.  Thus, both effects combine together reduces the FT values quickly with height and take a
 form proportional to both.  Following \citet{1999ApJ...514..441O}, equilibrium density in the solar atmosphere is given as
 
\begin{equation}
\rho = \rho_0 exp\left[-\frac{R_\odot}{H}\left(1-\frac{R_\odot}{r}\right)\right],
\label{eq:density}
\end{equation}

and approximate change in density due to wave propagation as

\begin{equation}
d\rho \sim \frac{R_\odot}{r} exp\left[-\frac{R_\odot}{2H}\left(1-\frac{R_\odot}{r}\right)\right],
\label{eq:cdensity}
\end{equation}

where $R_\odot$ is the solar radius, $H$ is the density scale height $\approx 61$~Mm
for $T=10^6$~K, and $r=R_\odot + h$ (h is height above solar limb). 

As a measurable quantity, intensity $I\propto \rho^2$, thus, the change in intensity due to wave propagation
will be obtained as

\begin{equation}
dI\propto 2\rho d\rho ~~\propto \frac{R_\odot}{r} exp\left[-\frac{3R_\odot}{2H}\left(1-\frac{R_\odot}{r}\right)\right].
\label{eq:dint}
\end{equation}

\begin{figure}[htbp]
\centering
\includegraphics[width=8cm]{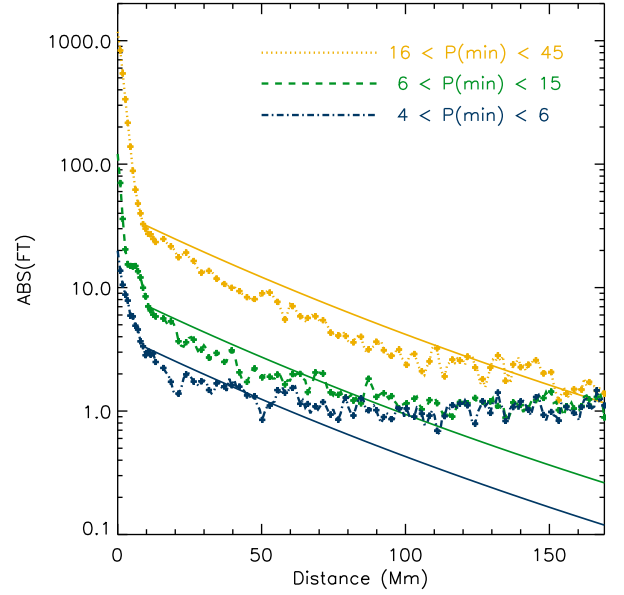}
\caption{Variation of Fourier transform (FT) values with height in different period (frequency) ranges as marked with different colours.
 Dotted lines indicate the observed decrease with height whereas continuous lines are those obtained after taking Eq.~\ref{eq:dint}
 into account.}
\label{fig:fit_pwr_ht}
\end{figure}

As Fourier transform~(FT) values depend upon amplitude of oscillations ($dI$) at different frequencies, 
in Fig.~\ref{fig:fit_pwr_ht} we show the decrease of FT values with height for the analysed plume region in different frequency ranges.
This variation of FT values obtained in different frequency ranges with height can be compared with the  expected decrease 
with height after taking the effect of stratification and area divergence  into account as obtained in Eq.~\ref{eq:dint}.

Upon choosing the reference FT value at very low height to obtain its decrease with height as per Eq.~\ref{eq:dint} in different frequency ranges,
 we found that the observed values were decreasing much more rapidly than those expected from Eq.~\ref{eq:dint}.
 Thus, we chose the reference value at a distance around $10$~Mm from the limb
 to obtain the curves according to Eq.~\ref{eq:dint} for all the frequency ranges. By comparison, we found that 
 the FT values in all the frequency ranges were still smaller than those predicted by Eq.~\ref{eq:dint}.
 FT values at farther higher heights flatten outs which indicates the appearance of white 
noise at those heights in the respective frequency ranges. 
Thus, from Fig.~\ref{fig:fit_pwr_ht}, it is clear that, in all  the frequency ranges, the FT values are decreasing more rapidly 
with height  than that expected from the stratification and area divergence together as per Eq.~\ref{eq:dint}. 
This may indicate the dissipation of slow waves with height due to various other parameters present.

The rapid decrease of FT values with height in comparison to Eq.~\ref{eq:dint} indicate the presence
 of additional physical components that result in the real dissipation of wave energy. These could be due to 
major damping parameters known such as thermal
conduction, compressive viscosity, radiation and other effects, or it could be due to a combination of several effects. 
Propagation and damping of slow magneto-acoustic waves in solar atmosphere are very well studied theoretically.
The models of \citet{2000A&A...362.1151N}, \citet{2004A&A...415..705D}, and \citet{2004ApJ...616.1232K}
 for coronal loops indicate thermal conduction to be the dominant dissipative mechanism among all.
\citet{1999ApJ...514..441O,2000ApJ...533.1071O} investigated the role of non-linear steepening of waves
 leading to dissipation via compressive viscosity  in polar plumes using 1-D and 2-D MHD codes.
 \citet{2002SoPh..209...89D} found that  thermal conduction could alone explain the damping of slow
 magneto-acoustic waves observed in coronal loops by TRACE. They also highlighted the relation between
 wave period (frequency) and damping length of slow waves \citep[Fig.~7a of ][]{2002SoPh..209...89D}.

Eq.~\ref{eq:dint} can be further simplified  using the geometric series approximation
$1/(1-x)=1+x+x^2+x^3+^{...}$ for $|x|<1$, here $x=-h/R_\odot$ with $h$ being very small as compared to $R_\odot$, and
thus, ignoring the higher order terms and taking $R_\odot/r\approx1$, we find

\begin{equation}
 dI\propto \frac{R_\odot}{r} exp\left[-\frac{3R_\odot}{2H}+\frac{3R_\odot^2}{2H(R_\odot+h)}\right] ~~\propto exp\left(-\frac{3h}{2H}\right).
 \label{eq:dintc}
\end{equation}

This is in agreement with the findings of \citet{2002SoPh..209...89D} and \citet{2004ApJ...616.1232K} for the observed
intensity perturbation of an acoustic wave propagating along the loop.

\begin{figure}[htbp]
\centering
\includegraphics[width=9cm]{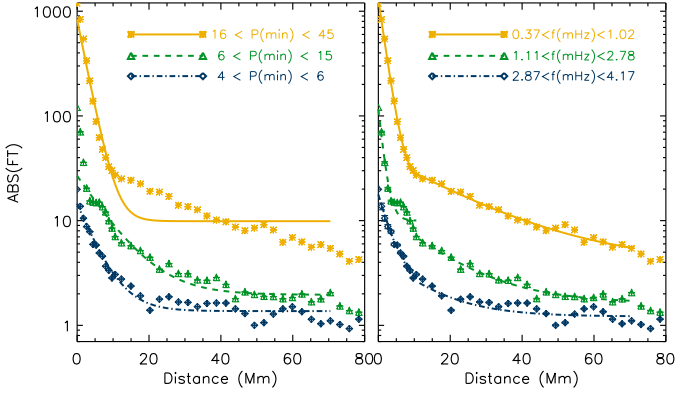}
\caption{Variation of Fourier transform (FT) values with height in different period (frequency) ranges as marked with different colours.
 Dotted lines are observed decrease with height whereas continuous lines are  fit applied to obtain dissipation
 lengths, as per Eq.~\ref{eq:damping}, using full height range (left panel), and in two height ranges (0--10 Mm and 10--70 Mm, right panel).
 The obtained dissipation lengths in two different height ranges are summarised in 
Tab.~\ref{tab:scale_ht}.}
\label{fig:damping_ht}
\end{figure}

Upon finding the evidence of true dissipation of slow waves, we can further obtain the dissipation length scale in this case. 
We thus add the effect of dissipation by multiplying the right-hand side of Eq.~\ref{eq:dintc} by $e^{-h/2H_d}$,
where $H_d$ is termed as \textquoteleft dissipation length\textquoteright\ for wave energy  as suggested by \citet{2004ApJ...616.1232K}.
We find

\begin{equation}
dI_d\propto exp\left(-\frac{3h}{2H}\right) exp\left({-\frac{h}{2H_d}}\right), 
\end{equation}

\begin{equation}
dI_d\approx A~exp\left(-\frac{3h}{2H}\right) exp\left({-\frac{h}{2H_d}}\right) +B.
\label{eq:damping}
\end{equation}

Where A and B are appropriate constants and $H\approx61$~Mm.
Thus, we obtained the dissipation length in different frequency ranges, by fitting the
FT values in different frequency ranges as per the Eq.~\ref{eq:damping}
using the MPFIT routines \citep{2009ASPC..411..251M}.
First, we fitted the FT values for heights up to 70 Mm in different frequency ranges and found that a single fit over whole height range
did not represent a good fit (left panel of Fig.~\ref{fig:damping_ht}). 
Therefore, as we found earlier that FT values were decreasing much more rapidly in the first 10~Mm distance,
 we divided the fit into two height ranges viz between 0--10 Mm and 10--70 Mm.
The resultant fits are presented in right panel of Fig.~\ref{fig:damping_ht} and dissipation lengths we obtained
in different period (frequency)  and height ranges are summarised in Tab.~\ref{tab:scale_ht}.

 Moreover, we also searched for any evidence of non-linearity in intensity oscillations at several heights. 
The relative amplitude of oscillations were about 8--10\% for the heights within first 10~Mm
 whereas that increased up to 13--15 \% beyond
10~Mm distance. These relative amplitudes of oscillations are within the limits of linearity, thus, 
dissipation of waves via compressive viscosity may not be the dominant mechanism here. 
  
Another possible dissipative mechanism that  may lead to dissipation of these waves could be thermal conduction.
\citet{2004ApJ...616.1232K} and \citet{2012A&A...546A..50K} found that dissipation length due to thermal conduction
is inversely proportional to the square of frequency (or directly proportional to the square of period) of wave propagation,
i.e. $H_d\propto 1/f^2$ (or $H_d\propto P^2$).
In our analysis, dissipation lengths obtained for wave propagation as summarised in Tab.~\ref{tab:scale_ht} 
indicate that there might exist some correlation between the dissipation lengths we obtained between 10-70~Mm height range 
and periods (frequencies) of wave propagation, and, more likely, it appears that  dissipation lengths are in direct (inverse) 
relation with periods (frequencies). No such relationship can be 
inferred from the dissipation lengths obtained at the lower height range (0-10~Mm).
Thus, the obtained results may indicate the presence of thermal conduction to be one of the dominant dissipative mechanism 
for the dissipation of slow waves at least in the height range of  10-70~Mm for the observed plume structure.
At the same, however, presence of other dissipative mechanisms showing relation between dissipation length and 
period (frequency) of wave propagation cannot be ruled out.

 Evidence of damping of slow magneto-acoustic waves were also found in the fan-like  
coronal loops by \citet{2012A&A...546A..50K} and suggested thermal conduction to be the dominant damping
 mechanism for the longer period slow magneto-acoustic waves. \citet{2011ApJ...734...81M}, however, found
 thermal conduction to be insufficient and magnetic field line divergence as the dominant factor in damping of
 slow waves with long-period oscillations in relatively cool and open structures of active regions. 
 Here, we found two distinct dissipation length scales of wave amplitude decay at two different height ranges (between 0--10 Mm and
 10--70 Mm) along the observed plume structure and the causes of those at lower heights are not clear.  
We would like to point out that damping lengths of slow wave propagation as a result of due to dissipation from
thermal conduction is of the order of 200--1000 Mm for the periods observed here 
\citep[see ][]{2002SoPh..209...89D,2011ApJ...734...81M}. 
The detection lengths found by \citet{2002SoPh..209...61D}, however, 
were of the order of 10~Mm for 5-min period propagating slow waves and those found by \citet{2009A&A...503L..25W}
were of the order of 70~Mm for the longer period waves. Clearly, damping lengths obtained from the forward modelling
 of thermal conduction are far too large to be able to explain the observed dissipation lengths of the order of a few Mm 
at different heights in the solar corona (see Tab.~\ref{tab:scale_ht}). 
A slightly enhanced thermal conductivity (four times of standard coronal value) can explain the smaller
values of damping length as suggested by \citet{2002SoPh..209...89D}. Thus, these observed dissipation lengths of 
slow waves may place a quantitative constraint on the theoretical models of slow wave propagation and dissipative
parameters in the solar corona.

\begin{table}[thbp]
\centering
\caption{Observed dissipation lengths of wave propagation in different period (frequency) ranges and at different heights.}
\begin{tabular}{cccc} 
\hline
 {\textbf{Period}} &  {\textbf{Frequency}} & \multicolumn{2}{c}{\textbf{Dissipation length (2\textit{H\textsubscript{d}})}} \\
 \textbf{(min)}                &  \textbf{(mHz)}           & \textbf{0--10 (Mm)} & \textbf{10--70 (Mm)} \\
\hline\hline
16--45 & 0.37--1.02 & 1.9 & 47.7     \\
6--15  & 1.11--2.78 & 1.3 & 22.2    \\
4--6    & 2.87--4.17 & 3.5 & 16.4     \\
\hline
\end{tabular}
\label{tab:scale_ht}
\end{table}

\subsection{Turbulence power spectra in the polar coronal hole}
\label{sec:turbulence}
\begin{figure*}[htbp]
\centering
\includegraphics[width=16cm]{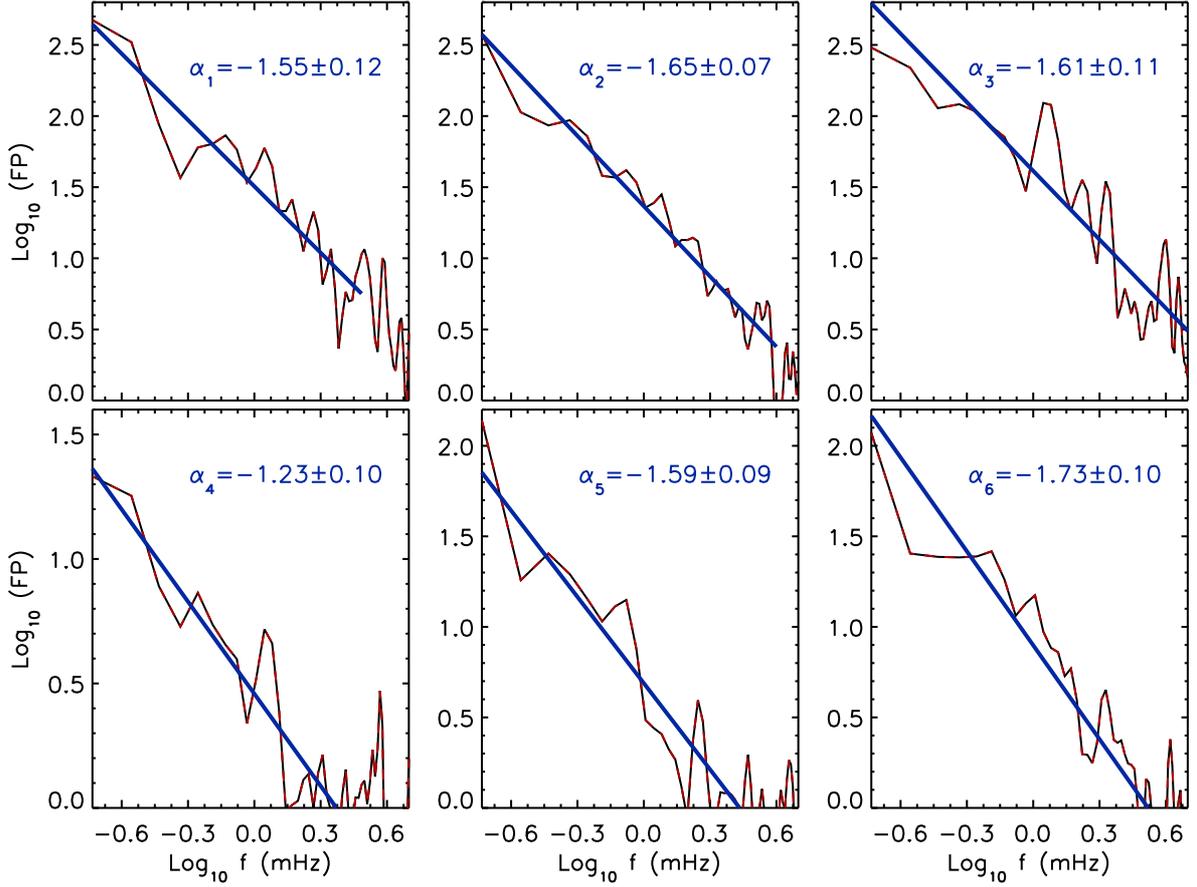}
\caption{Fourier power distribution with frequency for locations 1--6 marked in Fig.~\ref{fig:context}.
 Thin continuous lines represent Fourier powers obtained from original signal (after subtracting mean values). 
Powers were fitted with power-law distribution with spectral indices ($\alpha_n$) 
as labelled and are over-plotted with the thick continuous lines.}
\label{fig:fpwr}
\end{figure*}

From the various Fourier power maps and plots presented in the paper, we found that FT values at the base of the 
originating point of waves, and at further higher heights in the solar corona are different in various frequency ranges.
 Basically, the FT values at low frequencies were higher than those at high frequencies at similar heights,
 which made low frequency waves travel longer distances in the off-limb corona. 
In this section, we discuss the origin of such behaviour by analysing individual Fourier
 power spectra at several spatial locations in the polar coronal hole. 
  Fig.~\ref{fig:fpwr} shows the variation of Fourier power spectra with
 frequency at locations 1--6 in the polar coronal hole that are marked in Fig.~\ref{fig:context}. 
 The locations chosen close to the limb are about $30\arcsec$ away from the limb whereas those farther away
  are separated by about $35\arcsec$ in radial direction from the first location. 
We chose these selected lower height locations because the Fourier powers are present over the whole range of
 frequencies, from low to high ranges, only at these heights (see Fig.~\ref{fig:pwr_map}).
We chose different polar structures to perform our  analysis so as to demonstrate consistent
results in multiple structures.  
The power spectra we obtained are frequency dependent and have $1/f^\alpha$ dependence, where $\alpha$ is
the spectral power index. Individual power spectra at different locations were fitted with  
spectral indices, as labelled in the panels, and are summarised in Tab.~\ref{tab:distribution}. 
The values of $\alpha$\ are very close to $5/3$, and may indicate the presence of a Kolmogorov type power spectra
\citep[the nearly stationary regime of classical Kolmogorov turbulence;][]{1941DoSSR..30..301K} which is usually
observed in the solar wind turbulence \citep{2005LRSP....2....4B}.
These power spectra do not only show frequency dependence, but
 also show few small power peaks that resulted from oscillatory signatures due to waves propagation at those
 frequencies (power peaks are also observed in the wavelet plots, Fig.~\ref{fig:wvlt}).
 In our study, an analysis was performed over a three-hour time sequence, which resulted in frequencies in the range
 0.1~mHz to 40~mHz. These spectra fit very well for frequencies between $0.3$~mHz to $4.0$~mHz, and beyond that 
 white noise (flat spectra) start to appear depending upon the signal strength.
Because of this $1/f^\alpha$ dependence, power spectra are always dominated by low-frequency oscillations,
 which were used to be removed by trend subtraction in the earlier studies. Low-frequency dominance is independent
 of length of time sequence and exists over all the time-scales. The most appropriate example of this phenomena would
 be of solar cycle. If one takes the power spectrum (FT) of the raw time sequence of sunspot numbers over hundreds of years
  without any sophistication, the well-known 11-year period of sunspot numbers do not show up as sharp peak 
in the power spectrum \citep{1969WRR.....5..321M}. The peak gets smeared out in a strong power-law distribution
 which rises towards low frequencies \citep{1978ComAp...7..103P}.

The power spectra shown in Fig.~\ref{fig:fpwr} are generally called  \textquoteleft flicker noise/spectra\textquoteright\
 in astronomy \citep{1978ComAp...7..103P} and are  observed in interplanetary medium 
\citep[detailed in ][]{2005LRSP....2....4B,2010SSRv..156..135P} 
 in the frequency ranges mainly between $10^{-6}$~Hz to $10^{-3}$~Hz with different $\alpha$ values. 
 The $1/f^\alpha$ dependent power spectra are interpreted due to the presence of MHD turbulence 
\citep[see the review by][]{1995SSRv...73....1T}.
  Recently, \citet{2008ApJ...677L.137B} and \citet{2009ApJ...693.1022T} showed a similar kind of power spectra in
 $Ly\alpha$ observations from UVCS/SOHO in the polar coronal hole near the Sun at around 2~$R_\odot$ and in the outer
solar corona at 1.7~$R_\odot$, covering the fast and slow wind regions, respectively. 
\citet{2009ApJ...706..238T} found similar density fluctuations in slow and fast wind streams from
 the outer solar corona to  interplanetary space. 
Moreover, the photospheric magnetic field measured from MDI/SOHO also shows similar spectra \citep{2007ApJ...657L.121M}. 
\citet{2008ApJ...683L.207R} found the presence of a power-law distribution of oscillatory power with frequency in
the chromosphere with spectral indices ($\alpha$) $2.4$ and $2.5$ in the network and fibril regions, respectively.
We still need to perform more work to find the origin of $1/f^\alpha$ spectra either in  corona
 or elsewhere such as in photosphere or chromosphere.
 Height and structural dependence of spectral indices ($\alpha$) are still
unclear and detailed analysis needs to be performed to achieve any conclusive result. Here we report,
 however, the presence of $1/f^\alpha$\ spectra in near the Sun's polar coronal environment (within 1.1~$R_\odot$)
 from intensity ($\approx$density) fluctuations in time, which were reported earlier for higher heights
 \citep[e.g. ][]{2008ApJ...677L.137B,2009ApJ...693.1022T,2009ApJ...706..238T}.

\begin{figure*}[htbp]
\centering
\includegraphics[width=16cm]{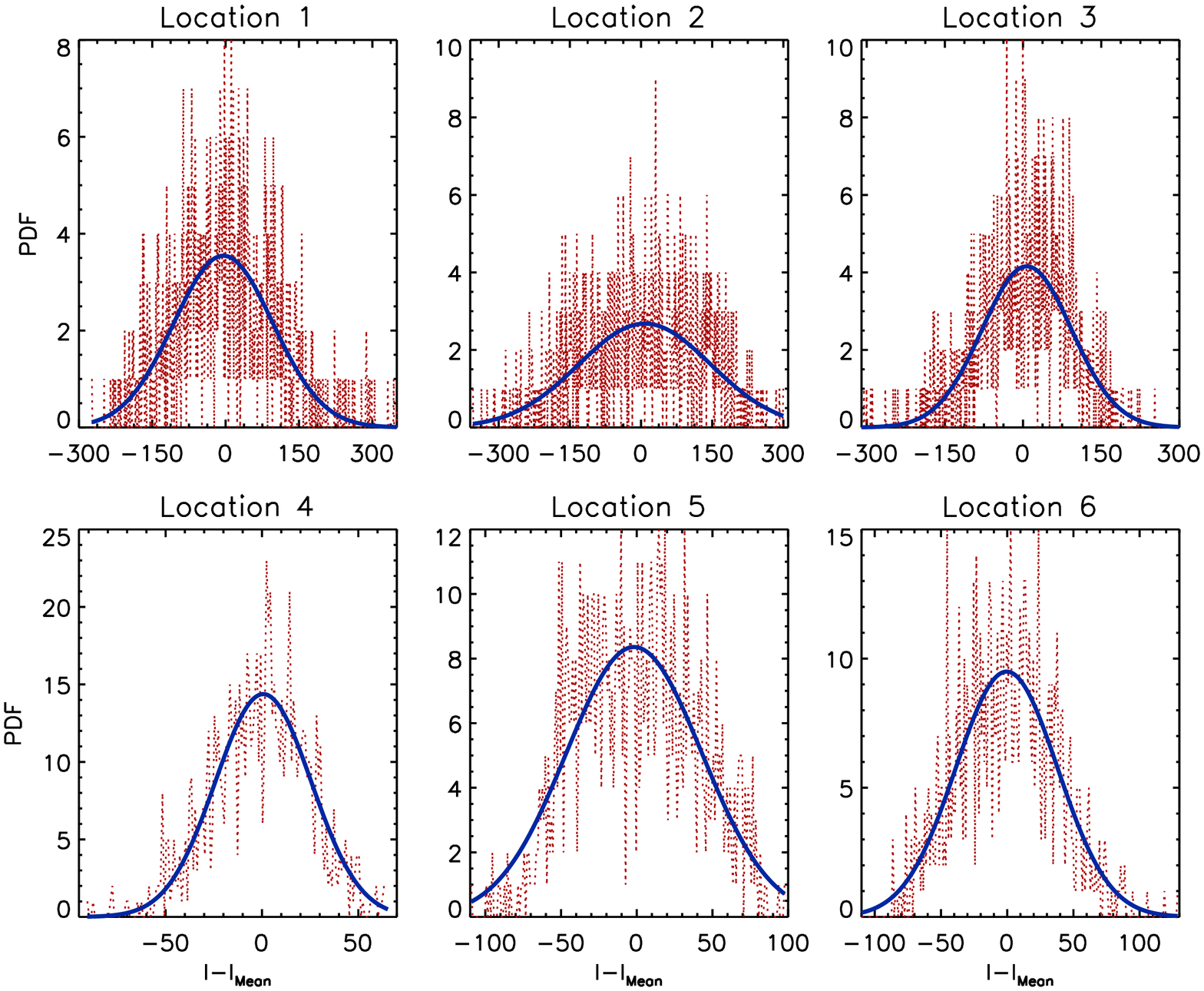}
\caption{PDFs obtained for intensity fluctuations  at 
 locations 1--6 marked in Fig.~\ref{fig:context}. PDFs were fitted with Gaussian distributions and are over-plotted
with continuous blue lines.}
\label{fig:idist}
\end{figure*}

Statistics of intensity fluctuations are important in the study of turbulence. 
The probability density function (PDF) of intensity fluctuations resulting from turbulence
are expected to have near-Gaussian distribution (not truly Gaussian), however, at small scales 
distribution can be highly non-Gaussian \citep[page 51,][]{1995tlan.book.....F}. 
Recently, \citet{2012ApJ...745..185L} performed the probability distribution analysis on intensity fluctuations
of solar prominence in space and time domain and found non-Gaussian and near-Gaussian distributions in both domains,
which were attributed to the presence of MHD turbulence. \citet{2008ApJ...683L.207R} found chromospheric
 PDFs to be non-Gaussian and photospheric PDFs to be nearly Gaussian in the quiet-Sun region, which were attributed to
intermittent phenomena associated with a turbulent cascade.   

To study nature of the distribution of intensity fluctuations in the polar coronal hole, 
we plot PDFs obtained at locations 1--6  in Fig.~\ref{fig:idist}. 
The mean value of intensity fluctuations were subtracted before obtaining the PDFs
at all locations. We fitted the PDFs with standard Gaussian function and  over-plotted for comparison.
 Gaussian profiles represent individual PDFs quite fairly. To verify any deviation from
Gaussanity, we estimate higher moments of intensity fluctuations. We measured the third and fourth moment of probability
distribution viz., skewness and kurtosis. Skewness is the measure of symmetry about the mean value whereas kurtosis is
that of peakedness. For a Gaussian distribution, skewness and kurtosis are zero and three, respectively.   
We calculated skewness and kurtosis for all the locations using  IDL routines $skewness.pro$ and 
$kurtosis.pro$ respectively, and the parameters we obtained are presented in Tab.~\ref{tab:distribution}
(IDL routine calculates excess kurtosis defined as \textit{\={k}~=~k}--3).
Fig.~\ref{fig:idist} and values presented in Tab.~\ref{tab:distribution} indicate that distribution of PDFs are 
near-Gaussian.  
 There could be, however,  uncertainties in intensity measurements due to data noise, thus, the observed
 distribution may be considered as nearly (or almost) Gaussian.  This departure from non-Gaussanity is expected as
 observations are integrated, line-of-sight intensity measurements. 
The line-of-sight measurements are still expected to capture qualitative features of turbulence
 such as power-law distribution and non-Gaussian fluctuations, however, the measurements may not lead to the similar theoretically 
predicted results. \citet{2012ApJ...745..185L} found power law distributed power spectra and similar 
near-Gaussian distribution of intensity fluctuations in the time domain from the observations of solar
 prominence where the quantitative signatures of turbulence were found.
Based on the results presented in Figs.~\ref{fig:fpwr} and \ref{fig:idist}, and Tab.~\ref{tab:distribution},
which confirm the power-law distribution of the Fourier power spectra and near-Gaussian distribution of intensity
 fluctuations at several locations, a presence of turbulence in the polar coronal hole may be inferred.

\begin{table*}[thbp]
\centering
\caption{Summary of distribution parameters at various locations.}
\begin{tabular}{ccccc} 
\hline
 &  &  & &\\
\raisebox{1.5ex}{\textbf{Location}} & \raisebox{1.5ex}{\textbf{Spectral Index ($\alpha$)}} &  \raisebox{1.5ex}{\textbf{Skewness ($s$)}} & \raisebox{1.5ex}{\textbf{Kurtosis (\={k})}} & \raisebox{1.5ex}{$\chi_{red}^{2}$} \\
\hline\hline
1 	&$1.55\pm0.12$	& $~0.29$	& $~0.17$	& 1.49 \\
2 	&$1.65\pm0.07$	& $-0.18$	& $-0.62$	& 1.48 \\
3 	&$1.61\pm0.11$	& $-0.31$	& $~0.31$	& 1.62 \\
4 	&$1.23\pm0.10$	& $-0.17$	& $~0.12$	& 6.27 \\
5 	&$1.59\pm0.09$	& $~0.06$	& $-0.70$	& 4.55 \\	
6 	&$1.73\pm0.10$	& $~0.08$	& $-0.22$	& 4.29 \\
\hline
\end{tabular}
\label{tab:distribution}
\end{table*}

Recently, \citet{2005ApJS..156..265C,2007ApJS..171..520C} and \citet{2010ApJ...708L.116V} developed
  models of self-consistent coronal heating and acceleration of fast solar wind. In these models,
 convective motions at the foot-points of magnetic flux
 tubes are assumed to generate wave-like fluctuations that propagate up into the extended corona,
 partially reflect back, develop into strong magnetohydrodynamic (MHD) turbulence, and then damped
 by an anisotropic turbulent cascade. In our study, we found the evidence of propagating MHD waves that were getting
heavily damped along with the evidence of turbulence. Although the damping mechanisms involved are not so clear,
  slow wave damping  due to thermal conduction, as suggested by \citet{2004ApJ...616.1232K}, appeared to be a more
 appropriate mechanism. \citet{2007A&A...473..931S} found turbulence to be a viable damping
mechanism to explain the spatial damping of $5-15$~min slow mode oscillations observed in quiescent limb prominences.
 Therefore, the role of turbulence as detected here may not be ignored in the damping of these waves.
Another model from \citet{2004ApJ...604..671F} suggested damping of MHD waves by background MHD turbulence.
 In this model, the waves propagating along magnetic field lines get distorted upon collisions with turbulent wave packets
 propagating in the opposite direction, resulting in a cascade to a smaller scale, and thus dissipation. 
These current models primarily deal with incompressible MHD waves and turbulence, however, the models highlight the importance
of MHD turbulence in wave damping. Thus, the obtained signatures of turbulence would be useful input in order to
understand the damping of waves even in the incompressible MHD framework. Nevertheless, to explain the current observations
 in detail, a model dealing with compressible MHD turbulence would be requested.

In this study,  we found propagating slow magneto-acoustic waves and the presence of turbulence in the polar coronal hole.
 Low-frequency waves were found to travel longer distances as compared to high-frequency waves, which were the result of
 the relatively high power content of the low-frequency waves. Turbulence spectra, which has similar energy
 distribution with respect to frequency, and were also found in our study, thus, may generate similar power 
distributed MHD waves. The origin and generation of these waves found in solar corona are, however, still unclear.
Recently, \citet{2013ApJ...775L..23W} modelled the propagating disturbances in fan-like coronal loops driven by 
up-flows with repetitive quasi-periodic tiny pulses generated by sporadic heating events (nanoflares) at the
 coronal base. The energy-frequency distribution of these events follows the power-law scaling with a broadband
 power-law spectrum. \citet{2013ApJ...775L..23W} found that the observed  propagating disturbances are mainly signatures of
wave propagation as the speeds are consistent with the wave speed. 
 Thus, the power-law distributed disturbances reported in this study may suggest that the 
 propagating disturbances are possibly driven by turbulent flows at the coronal base.
 Hence, turbulence spectra found in this study could also provide a viable mechanism for 
triggering of longer period waves in the solar atmosphere.

\section{Conclusion}
\label{sec:conclusion}

We presented the evidence of propagating slow magneto-acoustic waves in the polar coronal hole. The waves were reaching
very large distance in the off-limb region. We obtained a Fourier power map of the polar coronal hole in several frequency
ranges. We found that low-frequency waves were reaching very far out in the off-limb region whereas high-frequency
waves were confined to only a short distance in the off-limb region. We obtained wave dissipation lengths in different 
frequency and height ranges. We found that waves were getting heavily damped in the first 10~Mm  distance above
the limb with a dissipation length of  only a few Mm, whereas beyond that waves were damped slowly
with dissipation lengths of the order of several tens Mm, which also depends upon the frequencies.
Although the exact nature of  frequency dependence of dissipation lengths  is not clear, 
 their inverse relation may indicate the possibility of wave dissipation due to thermal conduction \citep{2004ApJ...616.1232K},
 however, the possibility of other dissipative mechanisms can not be ruled out.
Moreover, the cause of heavy damping (smaller dissipation lengths) of slow waves in the first 10~Mm distance is unclear.
The results we obtained may place a quantitative constraint on the theoretical models of slow wave propagation and dissipative
parameters in the solar corona. 

Individual Fourier power spectra obtained at several locations, indicated the clear power-law distribution, which 
may suggest the signature of turbulence in the polar coronal hole. We obtained probability density functions
 (PDFs) of intensity fluctuations at several locations and found those to be near-Gaussian distribution, and
suggested them arised due to the presence of turbulence. We report the presence of turbulence ($1/f^\alpha$)
 spectra in the near Sun polar coronal environment (within 1.1~$R_\odot$) from intensity fluctuations in the time domain,
 which were reported earlier for higher heights \citep{2008ApJ...677L.137B,2009ApJ...693.1022T}.
 The presence of turbulence could be important in excitation and damping of waves in the polar coronal hole
 \citep{1967SoPh....2..385S,2004ApJ...604..671F}.  
Because of power-law distribution, low-frequency waves always have high power content as compared to
high-frequency waves, and thus, are able to travel longer distances (longer detection lengths)
 comparatively. Spectral indices obtained
 from the power-law distribution indicates presence of Kolmogorov like turbulence \citep{1941DoSSR..30..301K}.
Results presented in this study may be important in the theoretical modelling of coronal heating
 and acceleration of the fast solar wind in the coronal holes from the MHD
 turbulence \citep{2007ApJS..171..520C,2010ApJ...708L.116V}.

\begin{acknowledgements}
The author is grateful to the referee for his/her valuable comments and clarifications, which
 improved the quality of presentation. The author thanks L. Teriaca and S. Solanki
for helpful discussions at the earlier stage of this work.
The author is thankful to D. Banerjee for his constant support throughout.
This work was supported through the INSPIRE Faculty Fellowship of the 
Department of Science and Technology (DST), Government of India.
The data used here are courtesy of NASA/SDO and AIA consortium.
\end{acknowledgements}

\bibliographystyle{aa.bst}
\bibliography{references}

\begin{thebibliography}{64}
\expandafter\ifx\csname natexlab\endcsname\relax\def\natexlab#1{#1}\fi

\bibitem[{{Banerjee} {et~al.}(2011){Banerjee}, {Gupta}, \&
  {Teriaca}}]{2011SSRv..158..267B}
{Banerjee}, D., {Gupta}, G.~R., \& {Teriaca}, L. 2011, \ssr, 158, 267

\bibitem[{{Banerjee} {et~al.}(2009){Banerjee}, {Teriaca}, {Gupta}, {Imada},
  {Stenborg}, \& {Solanki}}]{2009A&A...499L..29B}
{Banerjee}, D., {Teriaca}, L., {Gupta}, G.~R., {et~al.} 2009, \aap, 499, L29

\bibitem[{{Bemporad} {et~al.}(2008){Bemporad}, {Matthaeus}, \&
  {Poletto}}]{2008ApJ...677L.137B}
{Bemporad}, A., {Matthaeus}, W.~H., \& {Poletto}, G. 2008, \apjl, 677, L137

\bibitem[{{Berghmans} \& {Clette}(1999)}]{1999SoPh..186..207B}
{Berghmans}, D. \& {Clette}, F. 1999, \solphys, 186, 207

\bibitem[{{Bruno} \& {Carbone}(2005)}]{2005LRSP....2....4B}
{Bruno}, R. \& {Carbone}, V. 2005, Living Reviews in Solar Physics, 2, 4

\bibitem[{{Cranmer}(2007)}]{2007AIPC..932..327C}
{Cranmer}, S.~R. 2007, in American Institute of Physics Conference Series, Vol.
  932, Turbulence and Nonlinear Processes in Astrophysical Plasmas, ed.
  D.~{Shaikh} \& G.~P. {Zank}, 327--332

\bibitem[{{Cranmer}(2009)}]{2009LRSP....6....3C}
{Cranmer}, S.~R. 2009, Living Reviews in Solar Physics, 6, 3

\bibitem[{{Cranmer} \& {van Ballegooijen}(2005)}]{2005ApJS..156..265C}
{Cranmer}, S.~R. \& {van Ballegooijen}, A.~A. 2005, \apjs, 156, 265

\bibitem[{{Cranmer} {et~al.}(2007){Cranmer}, {van Ballegooijen}, \&
  {Edgar}}]{2007ApJS..171..520C}
{Cranmer}, S.~R., {van Ballegooijen}, A.~A., \& {Edgar}, R.~J. 2007, \apjs,
  171, 520

\bibitem[{{De Moortel}(2009)}]{2009SSRv..149...65D}
{De Moortel}, I. 2009, \ssr, 149, 65

\bibitem[{{De Moortel} \& {Hood}(2003)}]{2003A&A...408..755D}
{De Moortel}, I. \& {Hood}, A.~W. 2003, \aap, 408, 755

\bibitem[{{De Moortel} \& {Hood}(2004)}]{2004A&A...415..705D}
{De Moortel}, I. \& {Hood}, A.~W. 2004, \aap, 415, 705

\bibitem[{{De Moortel} {et~al.}(2004){De Moortel}, {Hood}, {Gerrard}, \&
  {Brooks}}]{2004A&A...425..741D}
{De Moortel}, I., {Hood}, A.~W., {Gerrard}, C.~L., \& {Brooks}, S.~J. 2004,
  \aap, 425, 741

\bibitem[{{De Moortel} {et~al.}(2002{\natexlab{a}}){De Moortel}, {Hood},
  {Ireland}, \& {Walsh}}]{2002SoPh..209...89D}
{De Moortel}, I., {Hood}, A.~W., {Ireland}, J., \& {Walsh}, R.~W.
  2002{\natexlab{a}}, \solphys, 209, 89

\bibitem[{{De Moortel} {et~al.}(2000){De Moortel}, {Ireland}, \&
  {Walsh}}]{2000A&A...355L..23D}
{De Moortel}, I., {Ireland}, J., \& {Walsh}, R.~W. 2000, \aap, 355, L23

\bibitem[{{De Moortel} {et~al.}(2002{\natexlab{b}}){De Moortel}, {Ireland},
  {Walsh}, \& {Hood}}]{2002SoPh..209...61D}
{De Moortel}, I., {Ireland}, J., {Walsh}, R.~W., \& {Hood}, A.~W.
  2002{\natexlab{b}}, \solphys, 209, 61

\bibitem[{{De Moortel} \& {Nakariakov}(2012)}]{2012RSPTA.370.3193D}
{De Moortel}, I. \& {Nakariakov}, V.~M. 2012, Royal Society of London
  Philosophical Transactions Series A, 370, 3193

\bibitem[{{DeForest} \& {Gurman}(1998)}]{1998ApJ...501L.217D}
{DeForest}, C.~E. \& {Gurman}, J.~B. 1998, \apjl, 501, L217

\bibitem[{{Farmer} \& {Goldreich}(2004)}]{2004ApJ...604..671F}
{Farmer}, A.~J. \& {Goldreich}, P. 2004, \apj, 604, 671

\bibitem[{{Frisch}(1995)}]{1995tlan.book.....F}
{Frisch}, U. 1995, {Turbulence. The legacy of A. N. Kolmogorov.}

\bibitem[{{Gupta} {et~al.}(2010){Gupta}, {Banerjee}, {Teriaca}, {Imada}, \&
  {Solanki}}]{2010ApJ...718...11G}
{Gupta}, G.~R., {Banerjee}, D., {Teriaca}, L., {Imada}, S., \& {Solanki}, S.
  2010, \apj, 718, 11

\bibitem[{{Gupta} {et~al.}(2009){Gupta}, {O'Shea}, {Banerjee}, {Popescu}, \&
  {Doyle}}]{2009A&A...493..251G}
{Gupta}, G.~R., {O'Shea}, E., {Banerjee}, D., {Popescu}, M., \& {Doyle}, J.~G.
  2009, \aap, 493, 251

\bibitem[{{Gupta} {et~al.}(2013){Gupta}, {Subramanian}, {Banerjee},
  {Madjarska}, \& {Doyle}}]{2013SoPh..282...67G}
{Gupta}, G.~R., {Subramanian}, S., {Banerjee}, D., {Madjarska}, M.~S., \&
  {Doyle}, J.~G. 2013, \solphys, 282, 67

\bibitem[{{Gupta} {et~al.}(2012){Gupta}, {Teriaca}, {Marsch}, {Solanki}, \&
  {Banerjee}}]{2012A&A...546A..93G}
{Gupta}, G.~R., {Teriaca}, L., {Marsch}, E., {Solanki}, S.~K., \& {Banerjee},
  D. 2012, \aap, 546, A93

\bibitem[{{Iroshnikov}(1964)}]{1964SvA.....7..566I}
{Iroshnikov}, P.~S. 1964, \sovast, 7, 566

\bibitem[{{Kiddie} {et~al.}(2012){Kiddie}, {De Moortel}, {Del Zanna},
  {McIntosh}, \& {Whittaker}}]{2012SoPh..279..427K}
{Kiddie}, G., {De Moortel}, I., {Del Zanna}, G., {McIntosh}, S.~W., \&
  {Whittaker}, I. 2012, \solphys, 279, 427

\bibitem[{{Klimchuk} {et~al.}(2004){Klimchuk}, {Tanner}, \& {De
  Moortel}}]{2004ApJ...616.1232K}
{Klimchuk}, J.~A., {Tanner}, S.~E.~M., \& {De Moortel}, I. 2004, \apj, 616,
  1232

\bibitem[{{Kolmogorov}(1941)}]{1941DoSSR..30..301K}
{Kolmogorov}, A. 1941, Akademiia Nauk SSSR Doklady, 30, 301

\bibitem[{{Kraichnan}(1965)}]{1965PhFl....8.1385K}
{Kraichnan}, R.~H. 1965, Physics of Fluids, 8, 1385

\bibitem[{{Krieger} {et~al.}(1973){Krieger}, {Timothy}, \&
  {Roelof}}]{1973SoPh...29..505K}
{Krieger}, A.~S., {Timothy}, A.~F., \& {Roelof}, E.~C. 1973, \solphys, 29, 505

\bibitem[{{Krishna Prasad} {et~al.}(2011){Krishna Prasad}, {Banerjee}, \&
  {Gupta}}]{2011A&A...528L...4K}
{Krishna Prasad}, S., {Banerjee}, D., \& {Gupta}, G.~R. 2011, \aap, 528, L4

\bibitem[{{Krishna Prasad} {et~al.}(2012){Krishna Prasad}, {Banerjee}, {Van
  Doorsselaere}, \& {Singh}}]{2012A&A...546A..50K}
{Krishna Prasad}, S., {Banerjee}, D., {Van Doorsselaere}, T., \& {Singh}, J.
  2012, \aap, 546, A50

\bibitem[{{Lemen} {et~al.}(2012){Lemen}, {Title}, {Akin}, {Boerner}, {Chou},
  {Drake}, {Duncan}, {Edwards}, {Friedlaender}, {Heyman}, {Hurlburt}, {Katz},
  {Kushner}, {Levay}, {Lindgren}, {Mathur}, {McFeaters}, {Mitchell}, {Rehse},
  {Schrijver}, {Springer}, {Stern}, {Tarbell}, {Wuelser}, {Wolfson}, {Yanari},
  {Bookbinder}, {Cheimets}, {Caldwell}, {Deluca}, {Gates}, {Golub}, {Park},
  {Podgorski}, {Bush}, {Scherrer}, {Gummin}, {Smith}, {Auker}, {Jerram},
  {Pool}, {Soufli}, {Windt}, {Beardsley}, {Clapp}, {Lang}, \&
  {Waltham}}]{2012SoPh..275...17L}
{Lemen}, J.~R., {Title}, A.~M., {Akin}, D.~J., {et~al.} 2012, \solphys, 275, 17

\bibitem[{{Leonardis} {et~al.}(2012){Leonardis}, {Chapman}, \&
  {Foullon}}]{2012ApJ...745..185L}
{Leonardis}, E., {Chapman}, S.~C., \& {Foullon}, C. 2012, \apj, 745, 185

\bibitem[{{Mandelbrot} \& {Wallis}(1969)}]{1969WRR.....5..321M}
{Mandelbrot}, B.~B. \& {Wallis}, J.~R. 1969, Water Resources Research, 5, 321

\bibitem[{{Markwardt}(2009)}]{2009ASPC..411..251M}
{Markwardt}, C.~B. 2009, in Astronomical Society of the Pacific Conference
  Series, Vol. 411, Astronomical Data Analysis Software and Systems XVIII, ed.
  D.~A. {Bohlender}, D.~{Durand}, \& P.~{Dowler}, 251

\bibitem[{{Marsh} {et~al.}(2011){Marsh}, {De Moortel}, \&
  {Walsh}}]{2011ApJ...734...81M}
{Marsh}, M.~S., {De Moortel}, I., \& {Walsh}, R.~W. 2011, \apj, 734, 81

\bibitem[{{Marsh} {et~al.}(2009){Marsh}, {Walsh}, \&
  {Plunkett}}]{2009ApJ...697.1674M}
{Marsh}, M.~S., {Walsh}, R.~W., \& {Plunkett}, S. 2009, \apj, 697, 1674

\bibitem[{{Matthaeus} {et~al.}(2007){Matthaeus}, {Breech}, {Dmitruk},
  {Bemporad}, {Poletto}, {Velli}, \& {Romoli}}]{2007ApJ...657L.121M}
{Matthaeus}, W.~H., {Breech}, B., {Dmitruk}, P., {et~al.} 2007, \apjl, 657,
  L121

\bibitem[{{Nakariakov} {et~al.}(2000){Nakariakov}, {Verwichte}, {Berghmans}, \&
  {Robbrecht}}]{2000A&A...362.1151N}
{Nakariakov}, V.~M., {Verwichte}, E., {Berghmans}, D., \& {Robbrecht}, E. 2000,
  \aap, 362, 1151

\bibitem[{{O'Dwyer} {et~al.}(2010){O'Dwyer}, {Del Zanna}, {Mason}, {Weber}, \&
  {Tripathi}}]{2010A&A...521A..21O}
{O'Dwyer}, B., {Del Zanna}, G., {Mason}, H.~E., {Weber}, M.~A., \& {Tripathi},
  D. 2010, \aap, 521, A21

\bibitem[{{Ofman} {et~al.}(1999){Ofman}, {Nakariakov}, \&
  {DeForest}}]{1999ApJ...514..441O}
{Ofman}, L., {Nakariakov}, V.~M., \& {DeForest}, C.~E. 1999, \apj, 514, 441

\bibitem[{{Ofman} {et~al.}(2000){Ofman}, {Nakariakov}, \&
  {Sehgal}}]{2000ApJ...533.1071O}
{Ofman}, L., {Nakariakov}, V.~M., \& {Sehgal}, N. 2000, \apj, 533, 1071

\bibitem[{{Ofman} {et~al.}(1997){Ofman}, {Romoli}, {Poletto}, {Noci}, \&
  {Kohl}}]{1997ApJ...491L.111O}
{Ofman}, L., {Romoli}, M., {Poletto}, G., {Noci}, G., \& {Kohl}, J.~L. 1997,
  \apjl, 491, L111+

\bibitem[{{O'Shea} {et~al.}(2006){O'Shea}, {Banerjee}, \&
  {Doyle}}]{2006A&A...452.1059O}
{O'Shea}, E., {Banerjee}, D., \& {Doyle}, J.~G. 2006, \aap, 452, 1059

\bibitem[{{O'Shea} {et~al.}(2007){O'Shea}, {Banerjee}, \&
  {Doyle}}]{2007A&A...463..713O}
{O'Shea}, E., {Banerjee}, D., \& {Doyle}, J.~G. 2007, \aap, 463, 713

\bibitem[{{Petrosyan} {et~al.}(2010){Petrosyan}, {Balogh}, {Goldstein},
  {L{\'e}orat}, {Marsch}, {Petrovay}, {Roberts}, {von Steiger}, \&
  {Vial}}]{2010SSRv..156..135P}
{Petrosyan}, A., {Balogh}, A., {Goldstein}, M.~L., {et~al.} 2010, \ssr, 156,
  135

\bibitem[{{Press}(1978)}]{1978ComAp...7..103P}
{Press}, W.~H. 1978, Comments on Astrophysics, 7, 103

\bibitem[{{Priest}(1984)}]{1984smh..book.....P}
{Priest}, E.~R. 1984, {Solar magneto-hydrodynamics}, ed. {Priest, E.~R.}

\bibitem[{{Reardon} {et~al.}(2008){Reardon}, {Lepreti}, {Carbone}, \&
  {Vecchio}}]{2008ApJ...683L.207R}
{Reardon}, K.~P., {Lepreti}, F., {Carbone}, V., \& {Vecchio}, A. 2008, \apjl,
  683, L207

\bibitem[{{Samadi} \& {Goupil}(2001)}]{2001A&A...370..136S}
{Samadi}, R. \& {Goupil}, M.-J. 2001, \aap, 370, 136

\bibitem[{{Samadi} {et~al.}(2001){Samadi}, {Goupil}, \&
  {Lebreton}}]{2001A&A...370..147S}
{Samadi}, R., {Goupil}, M.-J., \& {Lebreton}, Y. 2001, \aap, 370, 147

\bibitem[{{Singh} {et~al.}(2007){Singh}, {Dwivedi}, \&
  {Hasan}}]{2007A&A...473..931S}
{Singh}, K.~A.~P., {Dwivedi}, B.~N., \& {Hasan}, S.~S. 2007, \aap, 473, 931

\bibitem[{{Stein}(1967)}]{1967SoPh....2..385S}
{Stein}, R.~F. 1967, \solphys, 2, 385

\bibitem[{{Telloni} {et~al.}(2009{\natexlab{a}}){Telloni}, {Antonucci},
  {Bruno}, \& {D'Amicis}}]{2009ApJ...693.1022T}
{Telloni}, D., {Antonucci}, E., {Bruno}, R., \& {D'Amicis}, R.
  2009{\natexlab{a}}, \apj, 693, 1022

\bibitem[{{Telloni} {et~al.}(2009{\natexlab{b}}){Telloni}, {Bruno}, {Carbone},
  {Antonucci}, \& {D'Amicis}}]{2009ApJ...706..238T}
{Telloni}, D., {Bruno}, R., {Carbone}, V., {Antonucci}, E., \& {D'Amicis}, R.
  2009{\natexlab{b}}, \apj, 706, 238

\bibitem[{{Torrence} \& {Compo}(1998)}]{1998BAMS...79...61T}
{Torrence}, C. \& {Compo}, G.~P. 1998, Bull. Am. Meteorol. Soc., 79, 61

\bibitem[{{Tu} \& {Marsch}(1995)}]{1995SSRv...73....1T}
{Tu}, C. \& {Marsch}, E. 1995, Space Science Reviews, 73, 1

\bibitem[{{Uritsky} {et~al.}(2013){Uritsky}, {Davila}, {Viall}, \&
  {Ofman}}]{2013ApJ...778...26U}
{Uritsky}, V.~M., {Davila}, J.~M., {Viall}, N.~M., \& {Ofman}, L. 2013, \apj,
  778, 26

\bibitem[{{Verdini} {et~al.}(2010){Verdini}, {Velli}, {Matthaeus}, {Oughton},
  \& {Dmitruk}}]{2010ApJ...708L.116V}
{Verdini}, A., {Velli}, M., {Matthaeus}, W.~H., {Oughton}, S., \& {Dmitruk}, P.
  2010, \apjl, 708, L116

\bibitem[{{Verwichte} {et~al.}(2008){Verwichte}, {Haynes}, {Arber}, \&
  {Brady}}]{2008ApJ...685.1286V}
{Verwichte}, E., {Haynes}, M., {Arber}, T.~D., \& {Brady}, C.~S. 2008, \apj,
  685, 1286

\bibitem[{{Wang} {et~al.}(2013){Wang}, {Ofman}, \&
  {Davila}}]{2013ApJ...775L..23W}
{Wang}, T., {Ofman}, L., \& {Davila}, J.~M. 2013, \apjl, 775, L23

\bibitem[{{Wang} {et~al.}(2009){Wang}, {Ofman}, {Davila}, \&
  {Mariska}}]{2009A&A...503L..25W}
{Wang}, T.~J., {Ofman}, L., {Davila}, J.~M., \& {Mariska}, J.~T. 2009, \aap,
  503, L25

\bibitem[{{Yuan} \& {Nakariakov}(2012)}]{2012A&A...543A...9Y}
{Yuan}, D. \& {Nakariakov}, V.~M. 2012, \aap, 543, A9

\end{thebibliography}

\end{document}